# Canadian Publications in Library and Information Science: A Database of research by LIS academics and practitioners in Canada

Jean-Sébastien Sauvé[1], Madelaine Hare[2], Geoff Krause[3], Constance Poitras[4], Poppy Riddle[5]
Philippe Mongeon*[6]

[1] jean-sebastien.sauve@umontreal.ca
https://orcid.org/0000-0002-9472-2965
École de bibliothéconomie et des sciences de l'information, Université de Montréal

[2] maddie.hare@uottawa.ca
https://orcid.org/0000-0002-2123-9518
Digital Transformation and Innovation, University of Ottawa, Canada

[3] gkrause@dal.ca
https://orcid.org/0000-0001-7943-5119
Department of Information Science, Dalhousie University, Canada

[4] constance.poitras@umontreal.ca
https://orcid.org/0000-0002-3545-696X
École de bibliothéconomie et des sciences de l'information, Université de Montréal

[5] pnriddle@dal.ca
https://orcid.org/0000-0001-7862-7848
Department of Information Science, Dalhousie University, Canada

[6] PMongeon@dal.ca
https://orcid.org/0000-0003-1021-059X
Department of Information Science, Dalhousie University, Canada Centre interuniversitaire de recherche sur la science et la technologie (CIRST), Université du Québec à Montréal, Canada

* Corresponding author (pmongeon@dal.ca)

## Keywords

## Abstract

The aim of the Canadian publications in Library and Information Science (LIS) database is to help break down the silos in which the two main target audiences – LIS faculty members and academic librarians – conduct their research. As part of a larger project entitled "Breaking down research silos", we created a database of research contributions by Canadian LIS researchers (academics and practitioners). This was motivated by a desire to make research by Canadian LIS scholars and practitioners more visible and foster collaboration between these two groups. The aim of this paper is to introduce the database, describe the process through which it was created, provide descriptive statistics of the database content, and highlight areas for future development.

## Introduction

Library and Information Science (LIS) research in Canada has traditionally been the bailiwick of two groups: faculty members teaching in LIS departments, and academic librarians. While both groups are concerned with contributing to the development of professional theory and practice, performing research is a key aspect of university faculty members' workload. Academic librarians support research activities occurring at higher education institutions, and many are expected to devote part of their time to research activities as a part of their job descriptions (Ducas et al., 2020; Kandiuk & Sonne de Torrens, 2018). Quebec's francophone institutions are the exception, as academic librarians are not considered faculty members and do not share the same benefits (e.g., research sabbaticals, academic freedom) and research obligations as their colleagues from other provinces (Fox, 2007; Zavala Mora et al., 2023). The prioritization of scientific production in librarians' workload, however, is encouraged by professional associations, such as the Canadian Association of Research Libraries (CARL) and the Canadian Association of University Teachers (CAUT) (Babb, 2017). It follows that considering both LIS practitioners and academic research activities can help generate a more holistic understanding of these practices and of the contributions members of the LIS community make to the advancement of knowledge.

Several attempts to analyze the LIS research landscape in Canada have been made in the past decade (Paul-Hus et al., 2016; Julien & Fena, 2018; Shu & Mongeon, 2016; Mongeon et al., 2023). Many of these studies, partly because of their reliance on commercial databases with limited coverage (Mongeon & Paul-Hus, 2016), tend to overlook the contributions of librarians and particularly the French-speaking scientific community.

This paper introduces a dataset of publications authored by LIS academics and university librarians in Canada, which was created in the context of a research project exploring collaborations and interactions between the two groups. The dataset draws from sources like OpenAlex (Priem et al., 2022) and Google Scholar, which are open and more comprehensive than commonly used commercial databases like Web of Science and Scopus. We aim to increase the visibility of LIS research to better understand the diverse research areas and practices of the community and foster greater collaboration and engagement. In this paper, we describe the process through which the dataset was created, provide an overview of its contents, and highlight areas for future development the dataset and further research.

## Data and Methods

**General approach**

The objective of gathering research publications by two groups of people (academics and practitioners) implies a person-centred approach to the construction of our database, in which we gathered all the publications authored by a predetermined list of individuals as opposed to all the publications in a particular research area or a set of journals. The latter, publication-centred, approach would be more appropriate if the goal was to study a body of literature no matter who its contributors are. Accordingly, the process outlined below starts with the gathering of a list of individuals as the first step, and their research output (if any) as the second step.

This person-centred approach is in some regards less ambiguous than delineating the field based on topics. Selecting a set of individuals or organizational units to include in the database may not always be straightforward, but the boundaries between individuals and between organizational units tend to be more clear than disciplinary boundaries, especially in a field like LIS. Furthermore, due the multidisciplinary nature of the field, a topic-based approach to data collection would risk excluding research that sit at the periphery of what we might call the traditional or core LIS research topics. Similarly, our database would fail to capture the essence of a community of LIS researchers and practitioners if we considered all publications on information-related topics regardless of the authors' affiliations.

**Data collection and processing**

*List of academics and practitioners*

In summer 2022, we manually collected from the institutional websites the names of librarians from 93 Canadian universities and all researchers (including doctoral students and postdocs) of the eight Canadian organizational units (i.e., Faculty, Department, or School) offering an ALA-accredited program. For academic libraries, we collected a list of 93 Canadian universities and then consulted their websites to gather lists of academic librarians. Overall, 2,630 names (including duplicates, where individuals held multiple roles, or were affiliated with multiple institutions) were collected through this process, along with their institutional affiliation and status (academic or practitioner). Each person was also searched on Google Scholar and orcid.org, and the URLs of their profiles were recorded when found (620 Google Scholar profiles and 820 ORCID profiles).

*Google Scholar*

We used the scholar package in R to query the Google Scholar API and retrieve all the entries in the Google Scholar profiles of the 620 researchers for which a Google Scholar profile was found. In total, 23,176 publications were retrieved, linked to 572 Google Scholar profiles.

*ORCID*

Similarly, we used the ORCID API to retrieve the list of publications from the ORCID accounts we were able to identify. For the 820 ORCID profiles found for Canadian LIS researchers and practitioners, this stage yielded 4,938 publications linked to 204 distinct ORCID profiles. Note that ORCID profiles are managed by researchers themselves, and that listed publications are often linked via DOIs, ensuring higher data accuracy; this is offset by the ability of researchers to make their profiles and listed publications private, reducing the completeness of the available data.

*OpenAlex*

The full names of LIS researchers and practitioners were searched against OpenAlex authors, yielding a list of 154,847 (138,163 unique) potential author ID matches.

Publications records retrieved using ORCID containing DOIs were matched to OpenAlex works records using this as the identifier. Other works retrieved from ORCID profiles, as well as those from Google Scholar profiles, were matched to OpenAlex works by searching against the title field. OpenAlex author IDs linked to the works retrieved from Google Scholar profiles, as well as author IDs containing known ORCIDs were added to the list of potential author ID matches, bringing the total to 163,882 OpenAlex author IDs (139,466 unique). Works linked to these authors were then retrieved for manual disambiguation, alongside those previously retrieved from Google Scholar and ORCID that were not linked to OpenAlex works.

*Name disambiguation and verification*

Lists of practitioner/academic names and attributed works from ORCID, Google Scholar, and OpenAlex were supplied to our team, and were checked manually to determine whether these were the same individuals as our initial list.

This stage produced a list of 9,528 works attributed to 461 named individuals.

*Scopus*

Following the manual cleaning of publication lists, the list of linked authors was compared to the original list of LIS researchers and practitioners. Those not linked to any publications (2169) were searched for manually in Scopus. This yielded an additional 865 profiles and 4,247 publications, which were again matched to OpenAlex works using the DOI.

## Dataset overview

The result section provides a short overview of the dataset's content. An entity relationship diagram and a description of each field are available in Appendix A.

*Number of authors by group*

Overall, the dataset contains 1,326 distinct authors, 850 of which were classified as practitioners and 476 as academics. While we acknowledge that individuals can move from one group to the other or have dual roles at some point or for all their career, practitioners and academics are mutually exclusive categories in our dataset. Librarians teaching in LIS programs, for instance, were classified as practitioners in the dataset. It should also be noted that these statuses can change and that our dataset reflects imperfect information obtained in the summer of 2022.

*Number of records*

The dataset contains a total of 13,775 records out of which 8,230 are authored by at least one academic and 5,740 are authored by at least one practitioner. The number of records over for each group over time, presented in Figure 1, shows a peak in publications in 2021. This is caused by the fact that we conducted most of the data collection in 2022, and final steps (e.g., authors lookup in Scopus) were conducted in 2023 and 2024. Instead of dropping the 2022, 2023, and 2024 records from the dataset, we chose to include them and indicate in the dataset documentation a disclaimer that data from the 2022-2024 period is incomplete. Depending on the dataset usage, this may not be an issue. Furthermore, future attempts to update the dataset to include a complete publication record for 2022 onwards will be made easier by having some of the records already available.

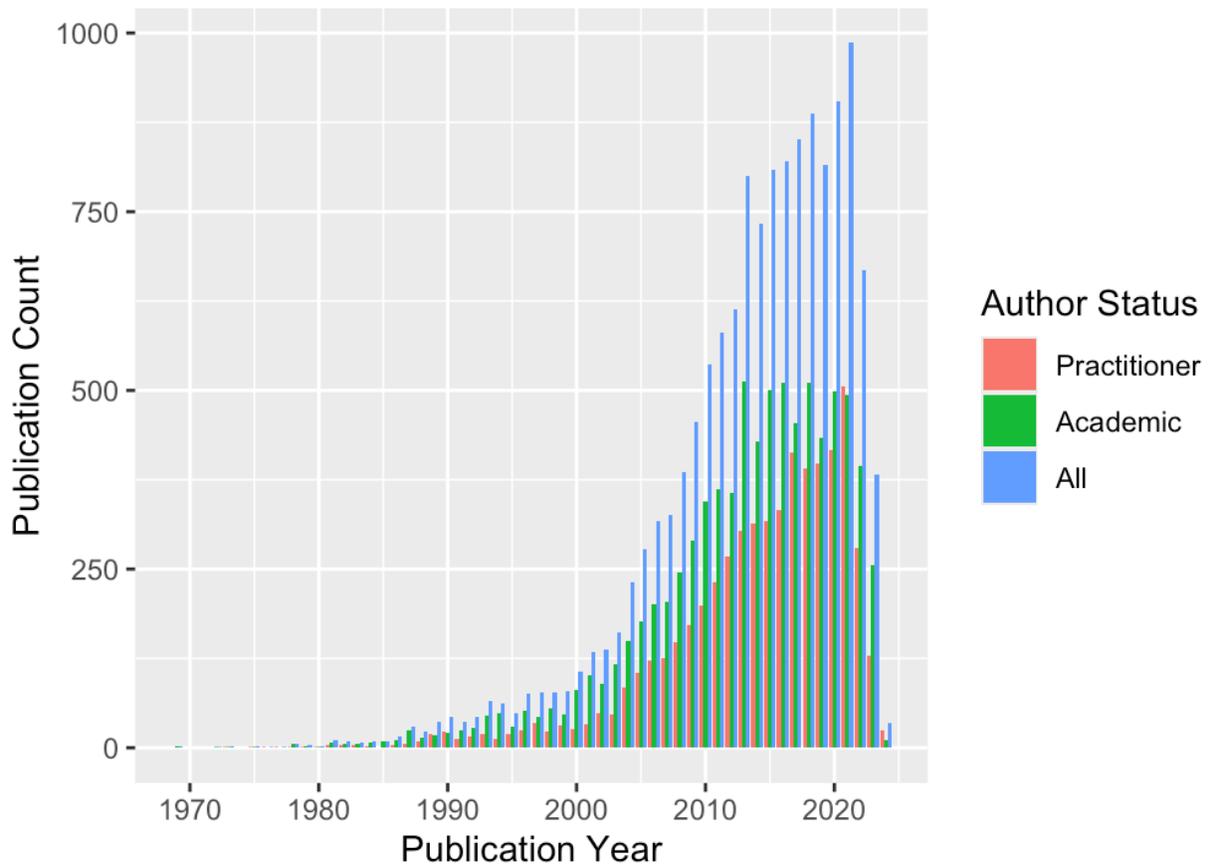

Figure 1. Yearly publication counts by author status

*Document types*

The dataset contains a wide array of document types, shown in Table 6, in part due to heterogeneous schemes used between source databases. Well over half are of type 'article', and over 90% of publications with an assigned type are of type 'article', 'conference paper', 'review', 'book chapter', or 'book review'. Over 14% of publications do not have a known type; in most cases the value is absent, but the number below includes a small amount marked explicitly as 'unknown' or 'other'. While not all document types necessarily represent research outputs, such a determination may be highly subjective, and we have opted not to filter these out, in order to allow users to tailor this to their needs.

Table 6. Number of records by type

| **Document type** | **Pract. pubs.** | **Acad. pubs.** | **All pubs.** |
|---|---:|---:|---:|
| All works | **5,740** | **8,230** | **13,775** |
| *Type unknown* | *1,362* | *652* | *1,968* |
| article | 2,580 | 4,934 | 7,413 |
| conference paper | 202 | 1,221 | 1,400 |
| review | 679 | 350 | 1,015 |
| book chapter | 122 | 682 | 798 |
| book review | 202 | 6 | 208 |
| editorial | 35 | 128 | 162 |
| book | 31 | 105 | 135 |
| report | 119 | 0 | 119 |
| conference presentation | 105 | 0 | 105 |
| note | 30 | 64 | 93 |
| research materials | 45 | 0 | 45 |
| dissertation/thesis | 47 | 1 | 48 |
| letter | 5 | 30 | 35 |
| preprint | 31 | 0 | 31 |
| monograph | 1 | 24 | 25 |
| protocol | 24 | 0 | 24 |
| erratum/correction | 6 | 16 | 22 |
| presentation | 18 | 0 | 18 |
| conference poster | 17 | 0 | 17 |
| meeting abstract | 15 | 0 | 15 |
| short survey | 3 | 9 | 12 |
| editorial material | 8 | 3 | 10 |
| *Other types (n<10 overall)* | *53* | *5* | *57* |

*Publication source*

Table 7 presents the top 20 most frequent publication sources (limited to journals and conferences) in the dataset as well as their rank and number of records for each group.

Table 7. Number of records by source (top 20 – Journals & Conferences)

| Source | All publications | | Practitioners | | Academics | |
|---|---|---|---|---|---|---|
| | Rank | N | Rank | N | Rank | N |
| Proceedings of the Annual Conference of CAIS / Actes du congrès annuel de l'ACSI | 1 | 206 | 10 | 38 | 1 | 191 |
| Proceedings of the American Society for Information Science and Technology | 2 | 184 | 28 | 13 | 2 | 175 |
| Evidence Based Library and Information Practice | 3 | 170 | 1 | 149 | 29 | 22 |
| The Deakin Review of Children's Literature | 4 | 132 | 2 | 131 | > 100 | 1 |
| Proceedings of the Association for Information Science and Technology | 5 | 124 | 78 | 6 | 3 | 119 |

| Journal | | | | | | |
|---|---|---|---|---|---|---|
| PLoS ONE | 6 | 101 | 11 | 33 | 8 | 68 |
| BMJ Open | 7 | 88 | 4 | 67 | 24 | 25 |
| Journal of the American Society for Information Science and Technology | 8 | 87 | > 100 | 3 | 4 | 85 |
| Documentation et bibliothèques | 9 | 82 | 59 | 8 | 6 | 74 |
| Partnership The Canadian Journal of Library and Information Practice and Research | 10 | 78 | 3 | 72 | 91 | 9 |
| Scientometrics | 11 | 77 | 59 | 8 | 7 | 70 |
| Journal of the Association for Information Science and Technology | 12 | 76 | > 100 | 3 | 5 | 75 |
| The Journal of Academic Librarianship | 12 | 76 | 6 | 59 | 29 | 22 |
| Library & Information Science Research | 14 | 75 | 44 | 10 | 8 | 68 |
| Journal of the Canadian Health Libraries Association / Journal de l'Association de bilbiothèques de la santé du Canada | 15 | 65 | 5 | 64 | > 100 | 3 |
| College & Research Libraries | 16 | 62 | 7 | 45 | 36 | 19 |
| Cataloging & Classification Quarterly | 16 | 62 | 24 | 14 | 12 | 50 |
| Journal of Documentation | 16 | 62 | > 100 | 4 | 10 | 61 |
| Education for Information | 19 | 56 | > 100 | 2 | 11 | 54 |
| Journal of the Medical Library Association JMLA | 20 | 55 | 8 | 41 | 43 | 17 |

## Conclusion

Ardanuy & Urbano (2017) commented on the weakening cooperation of LIS faculty and practitioners and cited an urgency to improve it "at a time when the discipline is at a crossroads of digital transformation that will require a commitment to research, development and innovation" (pg. 317). Making LIS publication data open and accessible may promote such cooperation, as it meets several objectives linked to the dissemination, promotion and preservation of LIS knowledge created by both academics and practitioners in Canada. Updating and improving the dataset on a continuing basis may contribute to improving visibility of and access to Canadian scientific contributions in the information sciences, highlighting the scientific contributions of librarians as researchers, encouraging further adoption of open data sharing practices, promoting inter-university and intersectoral exchanges between librarians and researchers, and advancing knowledge in the field.

# Acknowledgements

The authors acknowledge the support of the Social Sciences and Humanities Research Council of Canada, the Canadian Association of Research Libraries, the Maritime Institute for Science and Technology Studies, and the École de bibliothéconomie et des sciences de l'information of the Université de Montréal.

Les auteurs remercient le Conseil de recherches en sciences humaines du Canada, l'Association des bibliothèques de recherche du Canada, l'Institut maritime pour l'étude des sciences et technologies et l'École de bibliothéconomie et des sciences de l'information de l'Université de Montréal de leur soutien.

The authors extend their gratitude to Julia Crowell, Brandon Fitzgibbon, Catherine Gracey, Joanna Hiemstra, Vinson Li, Kydra Mayhew, Kellie Dalton, and Marc-André Simard for their contributions to the data collection.

# Author contributions

**J-S. S.** Conceptualization, Data curation, Funding acquisition, Investigation, Methodology, Project administration, Resources, Supervision, Validation, Writing – original draft, Writing – review & editing. **G. K.** Data curation, Formal Analysis, Investigation, Methodology, Validation, Visualization, Writing – original draft, Writing – review & editing. **P. R.** Data curation, Visualization, Writing – original draft, Writing – review & editing. **M. H.** Data curation, Writing – original draft, Writing – review & editing. **C. P.** Data curation, Writing – original draft, Writing – review & editing. **P. M.** Conceptualization, Data curation, Formal Analysis, Funding acquisition, Investigation, Methodology, Project administration, Resources, Supervision, Validation, Visualization, Writing – original draft, Writing – review & editing.

# Competing interests

The authors have no competing interests to declare.

# Data availability

The data set is available on Zenodo: https://doi.org/10.5281/zenodo.14302591.

# Appendix A. Database documentation

The entity relationship diagram of the dataset is presented in Figure 1.

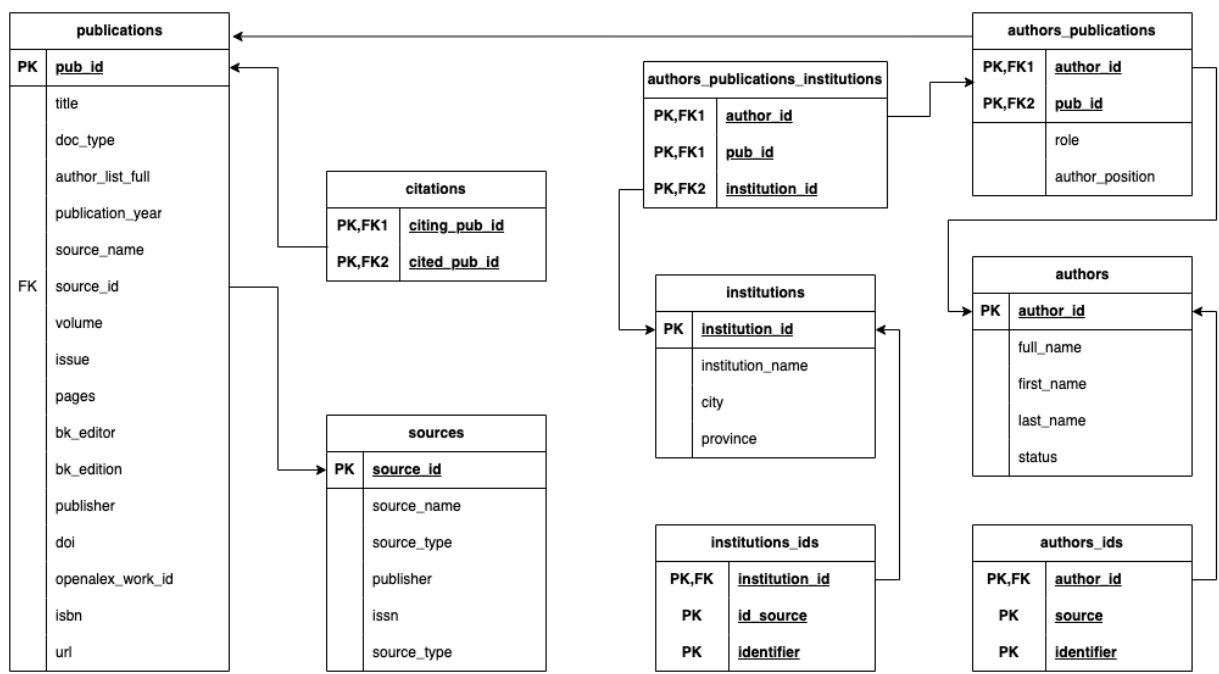

Figure 1. Entity relationship diagram

Table 1. Canadian LIS authors table (authors)

| Field | Description |
|---|---|
| author_id | Unique identifier for the publication in the LIS database |
| first_name | First name of author |
| last_name | Last name of author |
| full_name | Full name of author |
| status | Academic (Ph.D. student, a postdoctoral fellow, or a professor (assistant, associate, full, emeritus) in an organizational unit offering an ALA accredited degree) or practitioner (librarian position in a Canadian university) |

Table 2. Works table (publications)

| Field | Description |
|---|---|
| pub_id | Unique identifier for the publication in the LIS database |
| doi | Digital object identifiers |
| openalex_work_id | Identifier of the work in the OpenAlex database (URL format) |
| isbn | International standard book number (ISBN). |
| doc_type | Document type. Can take one of the following values: article; review; conference paper, book; edited book; book chapter. |
| publication_year | Year of publication |
| title | Title of the document |
| source_name | Title of the source (journal, conference, or book title for book chapters) |
| author_list_full | Full text listing of author names |
| volume | Volume number |
| issue | Issue number |
| pages | First and last pages separated by a hyphen. |
| bk_edition | Book edition |
| bk_editor | Name of book editor (for book chapters) |
| publisher | Publisher of the book/journal |
| source_id | Foreign key to the sources table |
| url | URL for the publication |

Table 3. Author publications table (authors_publications)

| Field | Description |
|---|---|
| author_id | Unique identifier for the author in the authors table |
| pub_id | Unique identifier for the work in the publications table |
| author_position | Position on the byline. |
| role | Role of the author on the work (author/editor) |

Table 4. Author IDs table (authors_ids)

| Field | Description |
|---|---|
| author_id | Unique identifier for the author in the authors table |
| source | Source for the identifier (e.g., OpenAlex, Scopus, Google Scholar, ORCID) |
| identifier | Identifier for the author in the source database |

Table 5. Publication source table (sources)

| Field | Description |
|---|---|
| source_id | Unique identifier for the source |
| source_name | Name of the source |
| publisher | Publisher name for the source |
| issn | ISSN for the source |
| source_type | OpenAlex source type (e.g., journal, conference) |

Table 6. Institutions table (institutions)

| Field | Description |
| --- | --- |
| institution_id | Unique identifier for the institution |
| institution_name | Name of the Canadian academic institution |
| city | Name of the city in which the institution is primarily located |
| province | Two-letter code of the province in which the institution is located |

Table 7. Institution IDs table (institutions_ids)

| Field | Description |
| --- | --- |
| institution_id | Unique identifier for the institution in the institutions table |
| id_source | Source database for the identifier (e.g., OpenAlex) |
| identifier | Identifier linked to the institution in the source database |

Table 8. Authorship institutional affiliation table (authors_publications_institutions)

| Field | Description |
| --- | --- |
| author_id | Author component of the authorship information in the authors_publications table |
| pub_id | Publication component of the authorship information in the authors_publications table |
| institution_id | Unique identifier for the affiliated institution in the institutions table |

Table 9. Citations table (citations)

| Field | Description |
| --- | --- |
| citing_pub_id | Unique identifier for the citing work in the publications table |
| cited_pub_id | Unique identifier for the cited work in the publications table |